\documentstyle[preprint,pra,eqsecnum,aps]{revtex}
\begin{document} \draft

\title{Coupled Harmonic Oscillators and Feynman's Rest of the
Universe}

\author{D. Han\footnote{electronic mail: han@trmm.gsfc.nasa.gov}}
\address{National Aeronautics and Space Administration, Goddard Space
Flight Center, Code 910.1, Greenbelt, Maryland 20771}

\author{Y. S. Kim\footnote{electronic mail: kim@umdhep.umd.edu}}
\address{Department of Physics, University of Maryland, College Park,
Maryland 20742}

\author{Marilyn E. Noz \footnote{electronic mail:
noz@nucmed.med.nyu.edu}}
\address{Department of Radiology, New York University, New York,
New York 10016}

\maketitle

\begin{abstract}

According to Feynman, the universe consists of two parts - the system
in which we are interested and the rest of the universe which our
measurement process does not reach.  Feynman then formulates the
density matrix in terms of the observable world and the rest of the
universe.  It is shown that coupled harmonic oscillators can serve as
an illustrative example for Feynman's ``rest of the universe.''  It is
pointed out that this simple example has far-reaching consequences in
many branches of physics, including statistical mechanics, measurement
theory, information theory, thermo-field dynamics, quantum optics, and
relativistic quantum mechanics.  It is shown that our ignorance of the
rest of the universe increases the uncertainty and entropy in the
system in which we are interested.

\end{abstract}

\narrowtext

\section{Introduction}\label{intro}
Because of its mathematical simplicity, the harmonic oscillator provides
soluble models in many branches of physics.  It often gives a clear
illustration of abstract ideas.  In his book on statistical
mechanics~\cite{fey72}, Feynman makes the following statement on the
density matrix. {\it When we solve a quantum-mechanical problem, what we
really do is divide the universe into two parts - the system in which we
are interested and the rest of the universe.  We then usually act as if
the system in which we are interested comprised the entire universe.
To motivate the use of density matrices, let us see what happens when we
include the part of the universe outside the system}.

The purpose of this paper is to study Feynman's rest of the universe and
related problems of current interest using a pair of coupled oscillators.
Starting from the classical mechanics of harmonic oscillators, we
formulate the symmetry of the oscillator system in terms of the group
Sp(2) which is of current interest and which is locally isomorphic to
the (2~+~1)-dimensional Lorentz group.  This symmetry is then extended
to the quantum mechanics of coupled oscillators.  This allows us to study
measurable and unmeasurable variables in terms of the two oscillator
coordinates.  The unmeasurable variable constitutes Feynman's rest of the
universe.  We shall study the effects of this unmeasurable variable to
the measurable variable in terms of the uncertainty relation and entropy.

In Sec.~\ref{classical}, we reformulate the classical mechanics of two
coupled oscillators in terms of the Sp(2) group.  The symmetry operations
include rotations and squeezes in the two-dimensional coordinate system
of two oscillator coordinates.  In Sec.~\ref{quantum}, this symmetry
property is extended to the quantum mechanics of the coupled oscillators.
In Sec.~\ref{wigf}, we use the Wigner phase-space distribution function
to see the effect of the unobservable variable on the uncertainty
relation in the observable world.  In Sec.~\ref{density}, we use the
density matrix to study the entropy of the system due to our ignorance
of the unobservable variable.  In Sec.~\ref{models}, it is shown that
the system of two coupled oscillators can serve as an analog computer
for many of the physical theories and models of current interest.
Section~\ref{concl} contains some concluding remarks.

\section{Coupled Oscillators in Classical Mechanics}\label{classical}
Two coupled harmonic oscillators serve many different purposes in
physics.  It is widely believed that this oscillator problem can be
formulated into a problem of quadratic equation of two variables, and
the quadratic equation can be separated by a simple rotation.  It is
true that the problem can be reduced to a quadratic equation, but it
is not true that this equation can be solved by one rotation.  Indeed,
in order to understand fully this simple problem, we have to employ
the SL(4,r) group which is isomorphic to the Lorentz group O(3,3)
with fifteen parameters~\cite{hkn95jm}.

In this paper, we do not need all the symmetries available from the
O(3,3) group.  We shall use one of the O(2,1)-like subgroups which is
by now a standard language in physics in all branches of physics.
Let us consider a system of coupled oscillators.  The Hamiltonian for
this system is
\begin{equation}\label{eq.1}
H = {1\over 2}\left\{{1\over m}_{1}p^{2}_{1} + {1\over m}_{2}p^{2}_{2}
+ A x^{2}_{1} + B x^{2}_{2} + C x_{1} x_{2} \right\}.
\end{equation}
By making scale changes of $x_{1}$ and $x_{2}$ to
$(m_{2}/m_{1})^{1/4} x_{1}$ and $(m_{1}/m_{2})^{1/4} x_{2}$
respectively, it is possible to write the above Hamiltonian in the
form~\cite{arav89,knp91}
\begin{equation}\label{eq.2}
H = {1\over 2m}\left\{p^{2}_{1} + p^{2}_{2} \right\} +
{1\over 2}\left\{A x_{1}^{2} + B x^{2}_{2} + C x_{1} x_{2} \right\} ,
\end{equation}
with $m = (m_{1}m_{2})^{1/2}$.   We can decouple this
Hamiltonian by making the coordinate transformation:

\begin{equation} \label{eq.3}
\pmatrix{y_{1} \cr y_{2}} = \pmatrix{\cos(\alpha/2) & -\sin(\alpha/2)
\cr \sin(\alpha /2) & \cos(\alpha /2)} \pmatrix{x_{1} \cr x_{2}}.
\end{equation}
Under this rotation, the kinetic energy portion of the Hamiltonian in
Eq.(\ref{eq.3}) remains invariant.  Thus we can achieve the decoupling
by diagonalizing the potential energy.  Indeed, the system becomes
diagonal if the angle $\alpha$ becomes
\begin{equation}\label{eq.4}
\tan \alpha  = {C\over B - A} .
\end{equation}
This diagonalization procedure is well known.  What is new in this note
is to introduce the new parameters $K$ and $\eta$ defined as
\begin{eqnarray}\label{eq.5}
K &=& \sqrt{AB - C^{2}/4} ,  \nonumber \\[.5ex]
\exp(-\eta) &=& \frac{A + B + \sqrt{(A - B)^{2} + C^{2}}}{4AB - C^{2}} .
\end{eqnarray}

In terms of this new set of variables, the Hamiltonian can be written as
\begin{equation}\label{eq.6}
H = {1\over 2m} \left\{p^{2}_{1} + p^{2}_{2} \right\} +
{K\over 2}\left\{e^{2\eta } y^{2}_{1} + e^{-2\eta } y^{2}_{2} \right\} ,
\end{equation}
with
\begin{eqnarray}\label{eq.7}
y_{1} &=& x_{1} \cos{\alpha \over 2} - x_{2}
\sin{\alpha \over 2} , \nonumber \\[0.5ex]
y_{2} &=& x_{1} \sin{\alpha \over 2} + x_{2} \cos{\alpha \over 2} .
\end{eqnarray}
In this way, we can study the symmetry properties of the coupled
oscillators systematically using group theoretical methods.  The coordinate
rotation of Eq.(\ref{eq.3}) is
generated by
\begin{equation}\label{eq.8}
J_{0} = - {i\over 2} \left\{x_{1} {\partial \over \partial x_{2}} - x_{2}
{\partial \over \partial x_{1}}\right\}.
\end{equation}
The $\eta$ variable changes the scale of $y_{1}$ in one
direction while changing that of $y_{2}$ in the opposite direction.  If
$y_{1}$ is expanded then $y_{2}$ becomes contracted so as to preserve the
product $y_{1}y_{2}$.  This is called the squeeze transformation.  The
squeeze operation which changes the scales of $y_{1}$ and $y_{2}$ is
generated by
\begin{equation}\label{eq.9}
S_{y} = -{i\over 2}\left\{y_{2} {\partial \over \partial
y_{2}} - y_{1} {\partial \over \partial y_{1}} \right\}.
\end{equation}
From the linear transformation of Eq.(\ref{eq.3}), this
expression can be written as
\begin{equation}\label{eq.10}
S_{y} = S_{1} \cos \alpha  - S_{2} \sin \alpha  ,
\end{equation}
with
\begin{eqnarray}\label{eq.11}
S_{1} &=& -{i\over 2}\left\{x_{1} {\partial \over \partial
x_{1}} - x_{2} {\partial \over \partial x_{2}} \right\}, \nonumber \\[.5ex]
S_{2} &=& -{i\over 2}\left\{x_{1} {\partial \over \partial x_{2}} +
x_{2} {\partial \over \partial x_{1}}\right\}.
\end{eqnarray}
Indeed, the generators $J_{0}, S_{1}$ and $S_{2}$ satisfy the commutation
relations:
\begin{equation}\label{eq.12}
[J_{0}, S_{1}] = iS_{2} ,\quad [J_{0}, S_{2}] = -iS_{1} ,
\quad [S_{1}, S_{2}] = iJ_{0} ,
\end{equation}
This set of commutation relations is identical to that for the
group SU(1,1) which is locally isomorphic to the (2 + 1)-dimensional
Lorentz group~\cite{knp91}.  If the problem is extended to the
four-dimensional phase space consisting of the $x_{1}, x_{2}, p_{1}$,
and $p_{2}$ variables, the symmetry group is Sp(4) which is locally
isomorphic to O(5,2)~\cite{knp91,dir63,hkn90}.  These groups are known
to provide the standard language for the one-mode and two-mode squeezed
states respectively~\cite{knp91}, in addition to their traditional
roles in other branches of physics, including classical mechanics,
nuclear, elementary particle and condensed matter physics.

\section{Quantum Mechanics of Coupled Oscillators}\label{quantum}
If $y_{1}$ and $y_{2}$ are measured in units of $(mK)^{1/4} $,
the ground-state wave function of this oscillator system is
\begin{equation}\label{eq.13}
\psi_{\eta}(x_{1},x_{2}) = {1 \over \sqrt{\pi}}
\exp{\left\{-{1\over 2}(e^{\eta} y^{2}_{1} + e^{-\eta} y^{2}_{2})
\right\} } .
\end{equation}
The wave function is separable in the $y_{1}$ and $y_{2}$ variables.
However, for the variables $x_{1}$ and $x_{2}$, the story is quite
different.

The key question is how the measurement or non-measurement of one
variable affects the world of the other variable.  If we are not able to
make any measurement in the $x_{2}$ space, how does this affect the quantum
mechanics in the $x_{1}$ space.  This effect is not trivial.  Indeed, the
$x_{2}$ space in this case corresponds to Feynman's rest of the universe,
if we only know how to do quantum mechanics in the $x_{1}$ space.  We shall
discuss in this paper how we can carry out a quantitative analysis
of Feynman's rest of the universe.

Let us write the wave function of Eq.(\ref{eq.13}) in terms of
$x_{1}$ and $x_{2}$, then
\begin{eqnarray}\label{eq.14}
\lefteqn{\psi_{\eta}(x_{1},x_{2}) = {1 \over \sqrt{\pi}}
\exp\left\{-{1\over 2}\left[e^{\eta}(x_{1}\cos{\alpha\over 2} -
x_{2} \sin{\alpha\over 2})^{2} \right.\right.} \nonumber \\[1ex]
\mbox{} & \mbox{} & \mbox{} \hspace{22mm} \left.\left.+
e^{-\eta}(x_{1}\sin{\alpha\over 2} +
x_{2}\cos{\alpha\over 2})^{2} \right] \right\} .
\end{eqnarray}
If $\eta = 0$, this wave function becomes
\begin{equation}\label{eq.15}
\psi_{0} (x_{1},x_{2}) = \frac{1}{\sqrt{\pi}} \exp{ \left\{- {1\over
2}(x^{2}_{1} + x^{2}_{2}) \right\}} .
\end{equation}
For other values of $\eta$, the wave function of Eq.(\ref{eq.14})
can be obtained from the above expression by a unitary transformation
generated by the operators given in Eq.(\ref{eq.8}) and Eq.(\ref{eq.11}).
We should then be able to write Eq.(\ref{eq.14}) as
\begin{equation}\label{eq.16}
\sum^{}_{m_{1}m_{2}}A_{m_{1}m_{2}}(\alpha, \eta) \phi_{m_{1}}(x_{1})
\phi_{m_{2}}(x_{2}) ,
\end{equation}
where $\phi_{m}(x)$ is the m$^{th}$ excited-state oscillator wave
function.  The coefficients $A_{m_{1}m_{2}}(\eta)$ satisfy the unitarity
condition
\begin{equation}\label{eq.17}
\sum^{}_{m_{1}m_{2}}|A_{m_{1}m_{2}}(\alpha, \eta)|^{2} = 1 .
\end{equation}
It is possible to carry out a similar expansion in the case of
excited states.

The question then is what lessons we can learn from the situation in
which we are not able to make measurements on the $x_{2}$ variable.
In order to study this problem, we use the density matrix and the
Wigner phase-space distribution function.

\section{Wigner Function and Uncertainty Relation}\label{wigf}
In his book, Feynman raises the issue of the rest of the universe in
connection with the density matrix.  Indeed, the density matrix plays
the essential role when we are not able to measure all the variables in
quantum mechanics~\cite{neu55,wiya63}.  In the present case, we assume
that we are not able to measure the $x_{2}$ coordinate.  It is often
more convenient to use the Wigner phase-space distribution function to
study the density matrix, especially when we want to study the
uncertainty products in detail~\cite{fey72,knp91}.

\widetext
For two coordinate variables, the Wigner function is defined
as~\cite{knp91}
\begin{eqnarray}\label{eq.18}
\lefteqn{W(x_{1},x_{2}; p_{1},p_{2}) = \left({1\over \pi} \right)^{2}
\int \exp\left\{-2i(p_{1}y_{1} + p_{2}y_{2})\right\}} \nonumber\\[1.0ex]
\mbox{ } & \mbox{ } & \mbox{ } \hspace{45mm}
\times \psi^{*}(x_{1} + y_{1}, x_{2} + y_{2})
\psi(x_{1} - y_{1}, x_{2} - y_{2}) dy_{1} dy_{2} .
\end{eqnarray}
The Wigner function corresponding to the wave function of
Eq.(\ref{eq.14}) is
\begin{eqnarray}\label{eq.19}
\lefteqn{W(x_{1},x_{2};p_{1},p_{2}) = \left(\frac{1}{\pi} \right)^{2}
\exp\left\{-e^{\eta }(x_{1} \cos{\alpha \over 2} -
x_{2} \sin{\alpha \over 2})^{2} - e^{-\eta }(x_{1} \sin{\alpha \over 2}
+ x_{2} \cos{\alpha \over 2})^{2} \right.}
\hspace{50mm} \nonumber \\[.5ex]
\mbox{} & \mbox{} & \mbox{} \left.
- e^{-\eta }(p_{1} \cos {\alpha \over 2}
- p_{2} \sin{\alpha \over 2})^{2} - e^{\eta }(p_{1} \sin{\alpha \over 2}
+ p_{2} \cos{\alpha \over 2})^{2}\right\} .
\end{eqnarray}
If we do not make observations in the $x_{2}p_{2}$ coordinates,
the Wigner function becomes
\begin{equation}\label{eq.20}
W(x_{1},p_{1}) =\int W(x_{1},x_{2};p_{1},p_{2}) dx_{2} dp_{2} .
\end{equation}
The evaluation of the integral leads to
\begin{eqnarray}\label{eq.21}
W(x_{1},x_{2};p_{1},p_{2}) &=& \left\{{1 \over \pi^{2}(1
+ \sinh^{2}\eta \sin^{2}\alpha)}\right\}^{1/2} \nonumber \\[1.0ex]
&{}& \times \exp\left\{-\left(\frac{x^{2}_{1}}{\cosh\eta -
\sin\eta \cos\alpha}
+ \frac{p^{2}_{1}}{\cosh\eta + \sin\eta \cos\alpha} \right) \right\} .
\end{eqnarray}
\narrowtext
This Wigner function gives an elliptic distribution in the phase
space of $x_{1}$ and $p_{1}$.  This distribution gives the uncertainty
product of
\begin{equation}\label{eq.22}
(\Delta x)^{2}(\Delta p)^{2} = {1\over 4}(1 + \sinh ^{2}\eta \sin
^{2}\alpha ) .
\end{equation}
This expression becomes 1/4 if the oscillator system becomes uncoupled
with $\alpha = 0$.  Because $x_{1}$ is coupled with $x_{2}$, our
ignorance about the $x_{2}$ coordinate, which in this case acts as
Feynman's rest of the universe, increases the uncertainty in the
$x_{1}$ world which, in Feynman's words, is the system in which we are
interested.

\section{Density Matrix and Entropy}\label{density}
Since the Wigner function is constructed from the density matrix, it is
straightforward to show~\cite{knp91}
\begin{equation}\label{eq.23}
Tr[\rho (x_{1} ,x_{1})] = \int W(x_{1},p_{1}) dx_{1} dp_{1} ,
\end{equation}
and
\begin{equation}\label{eq.24}
Tr[\rho^{2}] = 2 \pi \int W^{2}(x_{1},p_{1}) dx_{1} dp_{1} .
\end{equation}
If we compute these integrals, $Tr(\rho ) = 1$, as it should be
for all pure and mixed states.  On the other hand, $Tr(\rho^{2})$
becomes
\begin{equation}\label{eq.25}
Tr(\rho^{2}) = 1/(1 + \sinh^{2}\eta \sin^{2}\alpha)^{1/2} ,
\end{equation}
which is in general less than one.  This gives a measure of impurity
and also the degree of the effect of our ignorance on the system
in which we are interested.

Let us translate this into the language of the density matrix.  If both
$x_{1}$ and $x_{2}$ are measured, the density matrix is
\begin{equation}\label{eq.26}
\rho(x_{1},x_{2};x'_{1},x'_{2}) =
\psi(x_{1},x_{2})\psi ^{*}(x'_{1},x_{2}') .
\end{equation}
In terms of the expansion of the wave function given in
Eq.(\ref{eq.16}),
\begin{eqnarray}\label{eq.27}
\rho (x_{1},x_{2}&;&x'_{1},x'_{2}) = \sum^{}_{n_{1}n_{2}}
\sum^{}_{m_{1}m_{2}}A_{m_{1}m_{2}}(\alpha, \eta)
A^{*}_{n_{1}n_{2}}(\alpha,\eta) \nonumber \\[1.0ex]
&{}& \times \phi_{m_{1}}(x_{1})
\phi_{m_{2}}(x_{2}) \phi^{*}_{n_{1}}(x_{1}) \phi^{*}_{n_{2}}(x_{2}) .
\end{eqnarray}
If both variables are measured, this is a pure-state density matrix.

On the other hand, if we do not make a measurement in the $x_{2}$ space,
we have to construct the matrix $\rho (x_{1},x_{1}')$ by taking the
trace over the $x_{2}$ variable:
\begin{equation}\label{eq.28}
\rho (x_{1}, x_{1}') = \int \rho (x_{1},x_{2};x'_{1},x_{2}) dx_{2} .
\end{equation}
Then the density matrix $\rho (x_{1},x_{1}')$ takes the form
\begin{equation}\label{eq.29}
\rho (x_{1},x_{1}') = \sum^{}_{m,n} C_{mn}(\alpha,\eta)
\phi_{m}(x_{1}) \phi^{*}_{n}(x_{1}') ,
\end{equation}
with $C_{mn}(\alpha,\eta ) = \sum^{}_{k} A_{mk}(\alpha,\eta)
A^{*}_{nk}(\alpha,\eta )$.  The matrix $C_{mn}(\alpha,\eta)$ is also
called the density matrix.  The matrix $C_{mn}$ is Hermitian and can
therefore be diagonalized.  If the diagonal elements are $\rho_{m}$, the
entropy of the system is defined as~\cite{neu55,wiya63}
\begin{equation}\label{eq.30}
S = - \sum^{}_{m} \rho _{m} \ln (\rho _{m}) .
\end{equation}
The entropy is zero for a pure state, and increases as the system becomes
impure.  Like $Tr(\rho^{2})$, this quantity measures the effect of our
ignorance about the rest of the universe.

It is very important to realize that the above form of entropy can be
defined irrespective of whether or not the system is in thermal
equilibrium.  As soon as we have the entropy, we are tempted to introduce
the temperature.  This is not always right.  The temperature can only be
introduced into the system in thermal equilibrium.

\section{Physical Models}\label{models}
There are many physical models based on coupled harmonic oscillators,
such as the Lee model in quantum field theory~\cite{sss61}, the
Bogoliubov transformation in superconductivity~\cite{fewa71}, two-mode
squeezed states of light~\cite{dir63,hkn90,cav85}, the covariant
harmonic oscillator model for the parton picture~\cite{kim89}, and
models in molecular physics~\cite{iac91}.  There are also models of
current interest in which one of the variables is not observed,
including thermo-field dynamics~\cite{ume82}, two-mode squeezed
states~\cite{yupo87,ekn89}, the hadronic temperature~\cite{hkn89}, and
the Barnet-Phoenix version of information theory~\cite{baph91}.  They
are indeed the examples of Feynman's rest of universe.  In all of
these cases, the mixing angle $\alpha$ is $90^{o}$, and the mathematics
becomes much simpler.
\widetext
The Wigner function of Eq.(\ref{eq.19}) then becomes
\begin{eqnarray}\label{eq.31}
\lefteqn{W(x_{1},x_{2};p_{1},p_{2}) = \left(\frac{1}{\pi} \right)^{2}
\exp\left\{-{1\over 2}\left[e^{\eta}(x_{1} - x_{2})^{2} +
e^{-\eta}(x_{1} + x_{2})^{2}
\right. \right.} \hspace{70mm} \nonumber \\ [1.0ex]
\mbox{ } & \mbox{ } & \mbox{ } \left.\left. + e^{-\eta }
(p_{1} - p_{2})^{2} + e^{\eta }(p_{1} + p_{2})^{2} \right] \right\}.
\end{eqnarray}
This simple form of the Wigner function serves a starting point for
many of the theoretical models including some mentioned above.

If the mixing angle $\alpha$ is $90^{o}$, the density matrix also
takes a simple form.  The wave function of Eq.(\ref{eq.14}) becomes
\begin{equation}
\psi_{\eta}(x_{1},x_{2}) = {1 \over \sqrt{\pi}}
\exp\left\{- {1\over 4}\left[e^{\eta}(x_{1} - x_{2})^{2} +
e^{-\eta}(x_{1} + x_{2} )^{2} \right]\right\} .
\end{equation}
As was discussed in the literature for several different
purposes~\cite{knp91,kno79aj,knp86}, this wave function can be
expanded as
\begin{equation}\label{expan}
\psi_{\eta }(x_{1},x_{2}) = {1 \over \cosh\eta}\sum^{}_{k}
(\tanh \eta)^{k} \phi_{k}(x_{1}) \phi_{k}(x_{2}) .
\end{equation}
From this wave function, we can construct the pure-state density matrix
\begin{equation}
\rho_{\eta}(x_{1},x_{2};x_{1}',x_{2}')
= \psi_{\eta}(x_{1},x_{2})\psi_{\eta}(x_{1}',x_{2}') ,
\end{equation}
which satisfies the condition $\rho^{2} = \rho $:
\begin{equation}
\rho_{\eta}(x_{1},x_{2};x_{1}',x_{2}') =
\int \rho_{\eta}(x_{1},x_{2};x_{1}'',x_{2}'')
\rho_{\eta}(x_{1}'',x_{2}'';x_{1}',x_{2}') dx_{1}'' dx_{2}'' .
\end{equation}
If we are not able to make observations on the $x_{2}$, we should
take the trace of the $\rho$ matrix with respect to the $x_{2}$
variable.  Then the resulting density matrix is
\begin{equation}\label{integ}
\rho_{\eta}(x_{1},x_{1}') = \int \psi_{\eta}(x_{1},x_{2})
\left\{\psi_{\eta}(x_{1}',x_{2})\right\}^{*} dx_{2} .
\end{equation}
If we complete the integration over the $x_{2}$ variable,
\begin{equation}
\rho_{\eta}(x_{1},x_{1}') = \left({1\over \pi\cosh(2\eta)}\right)^{1/2}
\exp\left\{-{1\over 4}[(x_{1} + x_{1}')^{2}/\cosh(2\eta)
+ (x_{1} - x_{1}')^{2}\cosh 2\eta] \right\} .
\end{equation}
\narrowtext
The diagonal elements of the above density matrix is
\begin{equation}
\rho(x_{1},x_{1}) = \left({1\over \pi \cosh 2\eta} \right)^{1/2}
\exp \left(-x_{1}^{2}/\cosh 2\eta \right) .
\end{equation}
With this expression, we can confirm the property of the density
matrix: $Tr(\rho) = 1$.  As for the trace of $\rho^{2}$, we can
perform the integration
\begin{eqnarray}\label{trace2}
Tr\left(\rho^{2} \right) &=& \int \rho_{\eta}(x_{1},x_{1}')
\rho_{\eta}(x_{1}',x_{1}) dx'_{1}dx_{1} \nonumber \\[1.ex]
&=& \left({1 \over \cosh\eta}\right)^{2} ,
\end{eqnarray}
which is less than one for nonzero values of $\eta$.

The density matrix can also be calculated from the expansion of the
wave function given in Eq.(\ref{expan}).  If we perform the integral
of Eq.(\ref{integ}), the result is
\begin{equation}\label{dmat}
\rho_{\eta}(x_{1},x_{1}') = \left({1 \over \cosh\eta}\right)^{2}
\sum^{}_{k} (\tanh\eta)^{2k} \phi_{k}(x_{1})\phi^{*}_{k}(x_{1}') ,
\end{equation}
which leads to $Tr(\rho) = 1$.  It is also straight-forward to compute
the integral for to $Tr(\rho^{2})$.  The calculation leads to
\begin{equation}
Tr\left(\rho^{2} \right)
= (1/\cosh \eta)^{4} \sum^{}_{k} (\tanh\eta)^{4k} .
\end{equation}
The sum of this series is $(1/\cosh\eta)^{2}$, which is the same
as the result of Eq.(\ref{trace2}).

This is of course due to the fact that we are not making measurement
on the $x_{2}$ variable.  The standard way to measure this ignorance
is to calculate the entropy defined as \cite{wiya63}
\begin{equation}
S = - Tr\left(\rho \ln(\rho) \right) .
\end{equation}
If we use the density matrix given in Eq.(\ref{dmat}), the entropy
becomes
\begin{equation}
S = 2\left\{(\cosh\eta)^{2}\ln(\cosh\eta) -
(\sinh\eta)^{2}\ln(\sinh\eta)\right\} .
\end{equation}
This expression can be translated into a more familiar form if
we use the notation
\begin{equation}
\tanh\eta = \exp\left(-{\omega \over kT}\right) .
\end{equation}
The ratio $\omega/kT$ is a dimensionless variable.  In terms of
this variable, the entropy takes the form
\begin{equation}
S = \left({\omega \over kT}\right)
\frac{1}{\exp(\omega/kT) - 1}
- \ln\left[1 - \exp(-\omega/kT)\right] .
\end{equation}
This is the entropy for a system of harmonic oscillators in thermal
equilibrium.  Thus, for this oscillator system, we can relate our
ignorance to the temperature.

\section{Concluding Remarks}\label{concl}
It is interesting to note that Feynman's rest of the universe appears
as an increase in uncertainty and entropy in the system in which we are
interested.  In the case of coupled oscillators, the entropy allows us
to introduce the variable which can be associated with the temperature.
The density matrix is the pivotal instrument in evaluating the entropy.
At the same time, the Wigner function is convenient for evaluating the
uncertainty product.  We can see clearly from the Wigner function
how the ignorance or the increase in entropy increases the uncertainty
in measurement.

The major strength of the coupled oscillator problem is that its
classical mechanics is known to every physicist.  Not too well known
is the fact that this simple device has enough symmetries to serve as
an analog computer for many of the current problems in physics.
Indeed, this simple system can accommodate the symmetries contained
in O(3,3) which is the group of Lorentz transformations applicable to
three space-like and three time-like dimensions~\cite{hkn95jm}.
The group O(3,3) is has many interesting subgroups.  Many, if not
most, of the symmetry groups in physics are subgroups of this O(3,3)
group.

The authors are grateful to Leehwa Yeh for many helpful comments and
criticisms, and for pointing out errors in a number of equations in an
earlier version of the manuscript.

\end{document}